# Is research with qualitative data more prevalent and impactful now? Interviews, case studies, focus groups and ethnographies[1]

Mike Thelwall, Tamara Nevill, University of Wolverhampton.

Researchers, editors, educators and publishers need to understand the mix of research methods used in their field to guide decision making, with a current concern being that qualitative research is threatened by big data. Although there have been many studies of the prevalence of different methods within individual narrow fields, there have been no systematic studies across academia. In response, this article assesses the prevalence and citation impact of academic research 1996-2019 that reports one of four common methods to gather qualitative data: interviews; focus groups; case studies; ethnography. The results show that, with minor exceptions, the prevalence of qualitative data has increased, often substantially, since 1996. In addition, all 27 broad fields (as classified by Scopus) now publish some qualitative research, with interviewing being by far the most common approach. This suggest that qualitative methods teaching and should increase, and researchers, editors and publishers should be increasingly open to the value that qualitative data can bring.
**Keywords**: qualitative research methods; interviews; case studies; focus groups; ethnography.

## Introduction

Many scholars have claimed that qualitative research increased in prevalence or importance within social science and health research, during the 1980s (Alasuutari, 2010; Custard, 1998; Moon, Brewer, Januchowski-Hartley, et al., 2016; Ping, 2008; Richards, 2002), sometimes describing this as a "qualitative turn" (Fritz, 2014; Murray, & Chamberlain, 1998). No systematic evidence seems to have been presented to justify this claim or to track more recent trends, except within a few narrow fields (see below). More recently, some have argued that qualitative methods, such as the in-depth interview, are threatened by the opportunities provided by big data (Savage & Burrows, 2009). Thus, it is not clear whether qualitative methods are still increasing in prevalence, whether some types of qualitative data have different trajectories, and which fields exploit qualitative data most. Moreover, no prior study has systematically assessed whether qualitative research is as impactful as other types within academia. This information may help researchers and research methods educators when selecting research methods.

Qualitative research may be loosely thought of as that which analyses unstructured, non-numeric data, but it has many conflicting definitions. One much narrower synthesised defined a qualitative research method as, "an iterative process in which improved understanding to the scientific community is achieved by making new significant distinctions resulting from getting closer to the phenomenon studied" (Aspers & Corte, 2019). This definition highlights aspects that are common to many qualitative research methods, but that are not always necessary to investigate qualitative data. Common qualitative research methods include grounded theory (Strauss & Corbin, 1997), discourse analysis (Gee & Handford, 2013), interpretive phenomenological analysis (Larkin, Watts, & Clifton, 2006), narrative analysis (Cortazzi, 2014) and thematic analysis (Braun & Clarke, 2006), all of which have substantial internal variations.

---

[1] Thelwall, M. & Nevill, T. (in press). Is research with qualitative data more prevalent and impactful now? Interviews, case studies, focus groups and ethnographies. *Library & Information Science Research*.

Different types of data are analysed in qualitative research, including interviews, documents and observations. In this article the focus is on the method by which the data is obtained rather than the method by which it is analysed. This is a pragmatic decision because there are too many named qualitative methods to investigate individually. In-depth qualitative **interviews** can gain deep insights into an issue (Edwards & Holland, 2020). They also give a direct voice to non-researchers, allowing them to articulate their perspective in their own words and reducing the chance that the researcher imposes their own perspective or constrains the scope of discussion. **Focus groups** involve discussions with groups of people in the hope that interactions will help participants to produce more useful information (Morgan, 1996). In contrast, **case studies** analyse a phenomenon in a specific context using multiple sources of evidence (Mills, Durepos, & Wiebe, 2009). The phrase "case study" is also sometimes used to describe quantitative research but it is a named standard qualitative research method. **Ethnography** is concerned with investigating and describing cultural groups. It has various definitions, with some emphasising that the groups should be explained from their own perspective rather than fitted into pre-existing concepts, and others emphasising participant observation for evidence gathering (Atkinson, Coffey, Delamont, Lofland, & Lofland, 2001; Hammersley & Atkinson, 2019). Whichever definition is accepted, it is primarily a qualitative approach, with data sources likely to be observational and/or documentary, although some quantitative data may be included within holistic analysis. **Documentary analysis** involves analysing one or more documents, such as policy documents, business meeting minutes or social media posts (Shaw, Elston, & Abbott, 2004; Snelson, 2016). All these methods produce unstructured non-numeric data suitable for qualitative research, irrespective of the analysis method used.

## Problem statement

Given the importance or research method choice for making appropriate contributions to a field, it is important to understand the value and prevalence of common approaches in a field. The aim of this paper is to assess the prevalence and impact of research exploiting qualitative data throughout academia in recent years. Whilst previous studies have analysed individual fields, this study analyses all academic research fields. This article focuses on common named methods to obtain the types of data that is characteristic of much qualitative research: interviews, case studies, focus groups and ethnographies. Documentary analysis is excluded because it can also be a quantitative technique, and it is much less frequently named in academic documents (e.g., "ethnography" had 32,061 matches in Scopus but "documentary analysis" had only 2,704 in October 2020). The results of this part may inform researchers, editors, research methods courses and publishers about the current and likely importance of qualitative research in all broad fields.

Citation impact is also investigated. Citations may be poor indicators of the value of qualitative research if it has a predominantly applied focus, informing professions and applications rather than future research. Nevertheless, trends in citation impact may provide useful context about the value or uptake of qualitative research within academia. The following research questions drive the study.

- In which academic fields did the prevalence of interviews, case studies, focus groups and ethnography change between 1996 and 2019?

- Was the citation impact of interviews, case studies, focus groups and ethnography above or below average in any academic fields between 1996 and 2019?

## Literature review

This section reviews some of the conflicting pressures on qualitative research to increase or decrease.

**Reaction against the simplifications of quantitative research**: In fields where it is a minority activity, qualitative research may be necessary to ground a research field, guarding against the simplifications and surface analyses necessary for quantitative research (Engel & Pai, 2013; Garcia & Gluesing, 2013), for example by exploring the social realities underlying statistics (Alasuutari, 2010). This is also a reason for calls for mixed methods research, combining qualitative and quantitative approaches within a single study (Tashakkori & Teddlie, 1998).

**Easier access online**: Qualitative methods may be quicker to conduct as access to the internet becomes internationally widespread. For example, interviews can now take place online through Skype or email (Burns, 2010), saving on travel time and saving on cost compared to phone calls for international interviews. Online focus groups are also possible, and whilst a traditional ethnography might need to be conducted in person, virtual ethnographies (Hine, 2000) do not.

**The threat of big data.** The importance of qualitative research might decline if big data methods are able to get similarly deep insights with algorithmic methods (Savage & Burrows, 2009). Although there are some qualitative big data methods, it is primarily a quantitative tool. The combination of large quantities of socially relevant data, such as tweets or Facebook posts, and intelligent algorithms might be capable of generating deep insights into issues, but the extent to which these challenge qualitative research in academia is unclear.

**Reducing gender inequalities in academia**: Qualitative research methods are more likely to be used by female scholars than male scholars in India (Thelwall, Bailey, Makita, et al., 2019), the USA (Thelwall, Bailey, Tobin, & Bradshaw, 2019) and the UK (Thelwall, Abdoli, Lebiedziewicz, & Bailey, 2020), so the increasing share of female-authored research (e.g., Thelwall, 2020) may have generated a greater preference for qualitative methods in academia overall, a second order effect.

**The impact agenda**: In some countries, governments are incentivising research with measurable benefits to society (Gunn & Mintrom, 2016). In the UK, for example, the Research Excellence Framework impact case studies are financially well rewarded evidence-based narratives about how research units generate societal benefits (Watermeyer & Hedgecoe, 2016). Rewarding societal impacts may push research towards more applicable directions and outreach activities, which inevitably involve engaging with non-academics. This may generate new opportunities for people-based research for which qualitative methods may be used, another second order effect.

### Empirical evidence

The dominant research methods for a field can evolve substantially over time, as shown by an analysis of 100 years of an applied psychology journal (Cortina, Aguinis, & DeShon, 2017). Many empirical studies of small fields or journals have manually checked the type of research method in a sample of documents, usually finding an increase in the prevalence of qualitative research and sometimes also reporting the most prevalent type. Mixed methods research, incorporating both qualitative and quantitative approaches, can address the weaknesses of both and has increased in prevalence for funded US health research (Plano Clark, 2010).

Quantitative (60% in the third decade), qualitative (24% in the third decade), and mixed methods (6% in the third decade) research increased in prevalence within three major tourism journals from 1990 to 2017, with conceptual (i.e., theory-based: Xin, Tribe, & Chambers, 2013) research declining sharply from 43% to 10% (Nunkoo, Thelwall, Ladsawut, & Goolaup, 2020). Qualitative and mixed methods research have also been claimed to be replacing conceptual research in the social sciences generally (Alasuutari, 2010). In contrast from 2004 to 2016 in event management, hospitality and tourism journals, the proportions of these types of research did not change much, except that mixed methods research increased, to 19% (Draper, Young Thomas, & Fenich, 2018). Mixed methods research may also be increasing within library and information science, but is often not explicitly described as such (Granikov, Hong, Crist, & Pluye, 2020), and quantitative research dominates the field (Ullah & Ameen, 2018). Thus, any increases in qualitative research may be partly as a replacement for conceptual research and partly as a component of mixed methods approaches rather than as a replacement for quantitative research.

Qualitative research also increased in prevalence and quality in selected management journals 1999-2008 (Bluhm, Harman, Lee, & Mitchell, 2011). Qualitative methods (except historical research) moderately increased in prevalence for empirical research in library and information science journals between 1985 and 2005 (Hider, & Pymm, 2008) and increased for mobile communication studies articles 1999-2012 (Taipale & Fortunati, 2014). In contrast, two studies have not found clear increases in the prevalence of qualitative research. There was little change in the share of qualitative research in leading small business and enterprise journals between 2001 and 2008 (Mullen, Budeva, & Doney, 2009). Qualitative methods increased their share in two higher education research journals from 1996–2000 to 2006–2010, but decreased their share in another (Wells, Kolek, Williams, & Saunders, 2015).

Some studies on the prevalence of qualitative methods have also differentiated between the different types, usually finding interviews to be the most common. Qualitative research increased from 2010–2012 to 2015–2017 in sports and exercise psychology journals, but formed a minority (18%), with interviews (85%) being the most common qualitative approach (McGannon, Smith, Kendellen, & Gonsalves, 2019). The amount of qualitative research related to surgery also increased to 2015, with interviews being most common (82%), usually of patients (64%), although only 8% were published in surgery journals (Maragh-Bass, Appelson, Changoor et al., 2016).

Case studies doubled in prevalence between 1961 and 2015 to about 18% in International Journal of Production Research (Manikas, Boyd, Pang, & Guan, 2019). Qualitative case studies also increased in prevalence in operations management journals 1992-2007, but many published studies had reporting limitations (Barratt, Choi, & Li, 2011).

No studies seem to have compared the impact of qualitative and quantitative research, but an analysis of Finnish and Danish sociology found that qualitative research was increasingly published in national journals (Erola, Reimer, Räsänen, & Kropp, 2015). If this is true for most other countries, then qualitative research would presumably be less cited as a result. In contrast, a study of 299 library and information science articles published from 2003 to 2013 found no difference in average citation impact between research using four qualitative methods and other articles (Jamali, 2018).

# Methods

The research design was to calculate the prevalence and citation impact of interviews, case studies, focus groups and ethnography in each Scopus broad field between 1996 and 2019. These four methods are qualitative, common and relatively straightforward to identify by keyword queries. Scopus was chosen as the database because it has wider coverage than the Web of Science, and particularly for non-English articles (Mongeon & Paul-Hus, 2016). Thus, Scopus may capture a larger share of qualitative research if it is frequently published in national journals (Alasuutari, 2010). There are 27 Scopus broad fields, which are mainly applied to whole journals. Journals are usually classified into multiple fields, so it is a relatively crude classification for articles but nevertheless gives a transparent method of roughly dividing articles into a manageable number of fields (Klavans & Boyack, 2017). The year 1996 was chosen as the starting point because Scopus expanded in that year so evidence from prior years would give unreliable trends. Only standard journal articles were included (e.g., excluding reviews and editorials) to give a consistent document type to analyse.

The titles, abstracts and author keywords of each Scopus article 1996-2019 were queried for the name of each research method in both singular and plural forms. Articles without an abstract or with a very short abstract (under 500 characters) were excluded because methods details are less likely to be mentioned in perfunctory or absent abstracts.

The citation impact of each article was calculated by log-transforming it with $\log(1+x)$ to reduce skewing, then dividing it by the average of the log-transformed citation counts of all article in the same field and year. For articles classified into multiple fields, the divisor was the average of all relevant field/year averages. The resulting figure is greater than 1 only if the article has a log-transformed citation count that is above average for its field and year. Averaging these normalised scores produces the Mean Normalised Log-transformed Citation Score (MNLCS) (Thelwall, 2017), which can reasonably be compared between fields and years.

# Results

## *Prevalence*

All four methods varied widely in prevalence between broad fields (Figure 1). All methods also increased in prevalence in all broad fields from 1996 to 2019, except that case studies decreased in Nursing. In most cases the method was at least twice as commonly mentioned in 2019 compared to 1996. Some notable exceptions to this rule are that case study research only increased marginally in Psychology, Health Professions and Neuroscience. Ethnography also increased only marginally in Nursing. The huge increases in focus group research is particularly remarkable, at least quadrupling in several fields, including its most used field, Nursing.

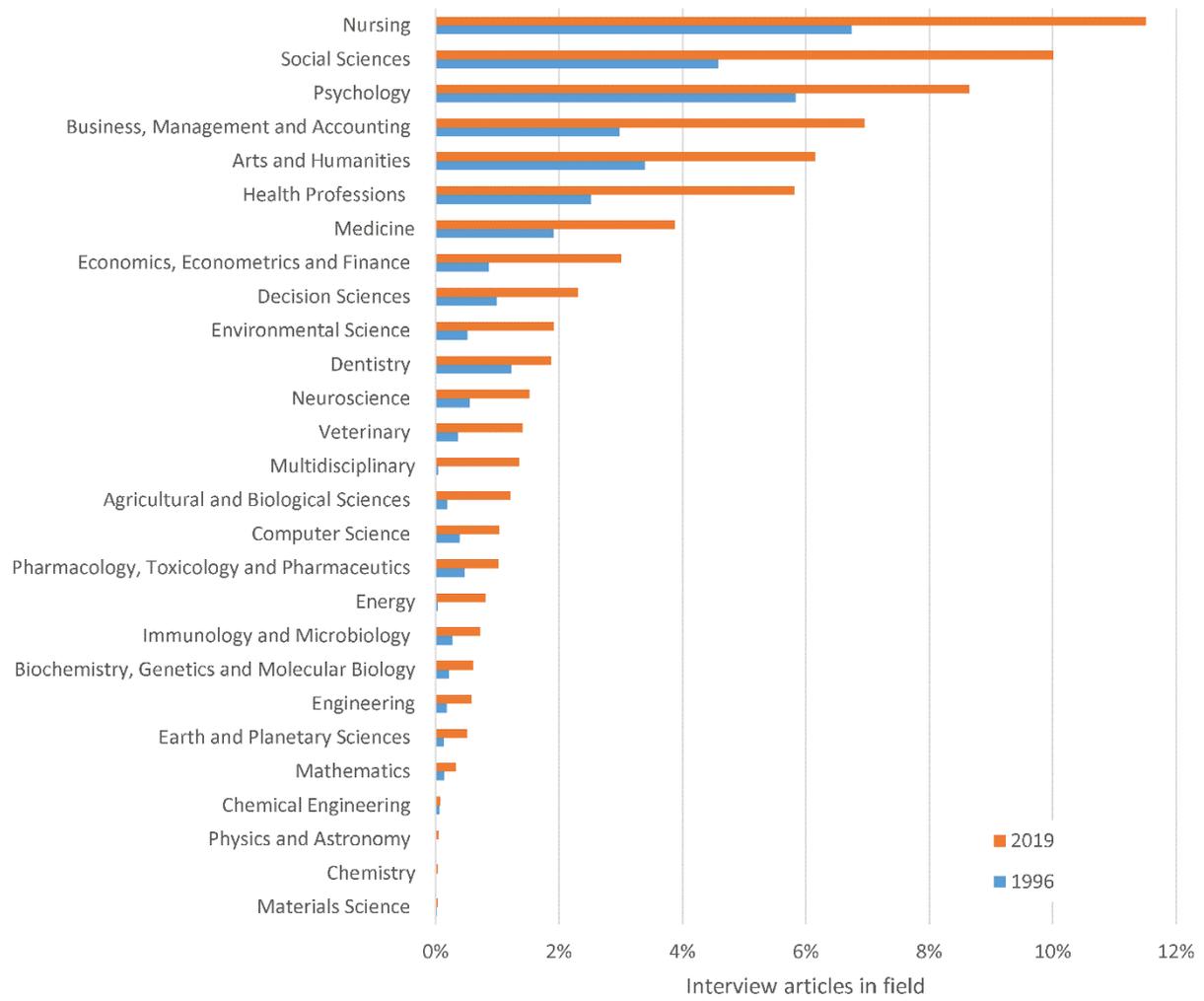

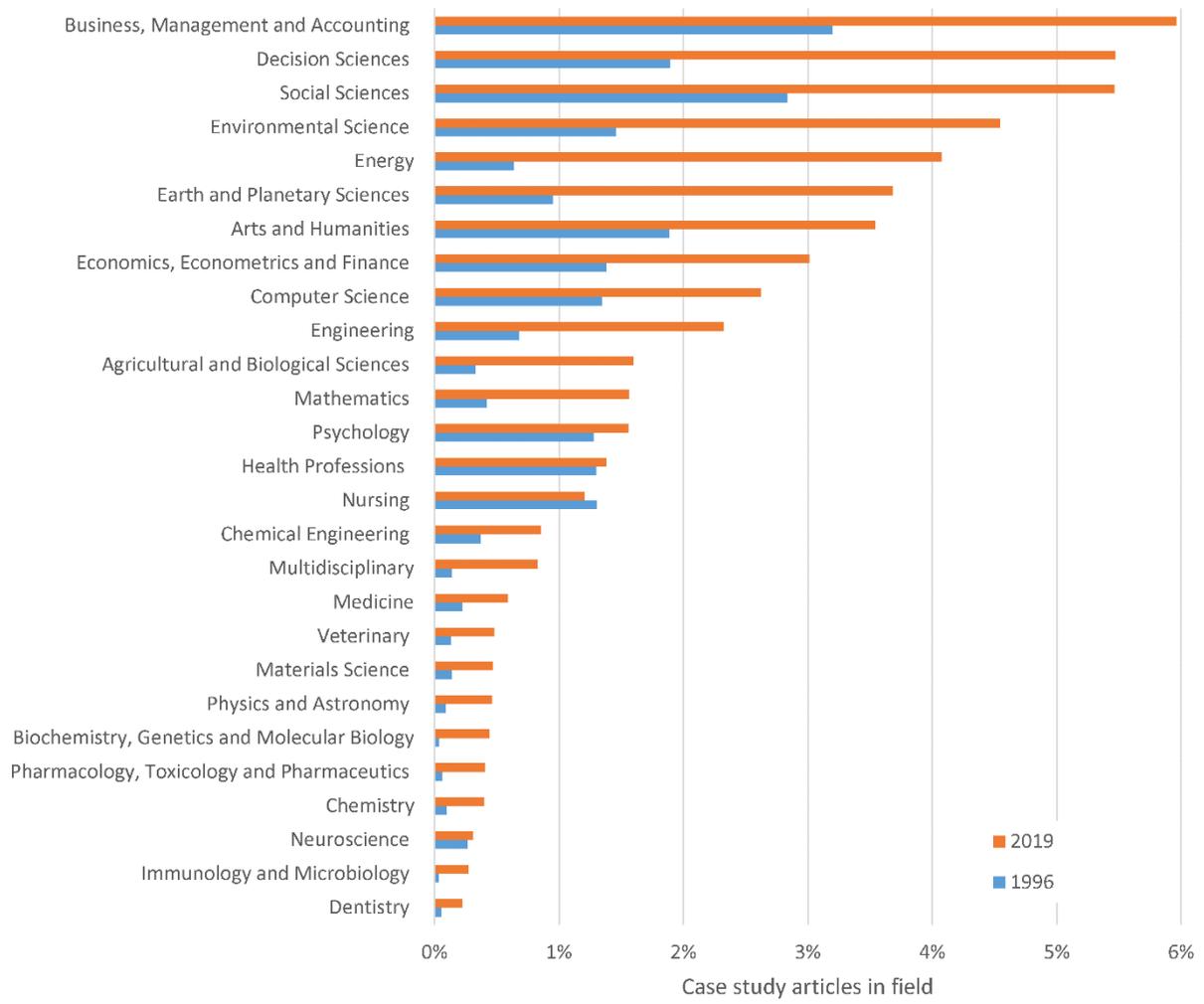

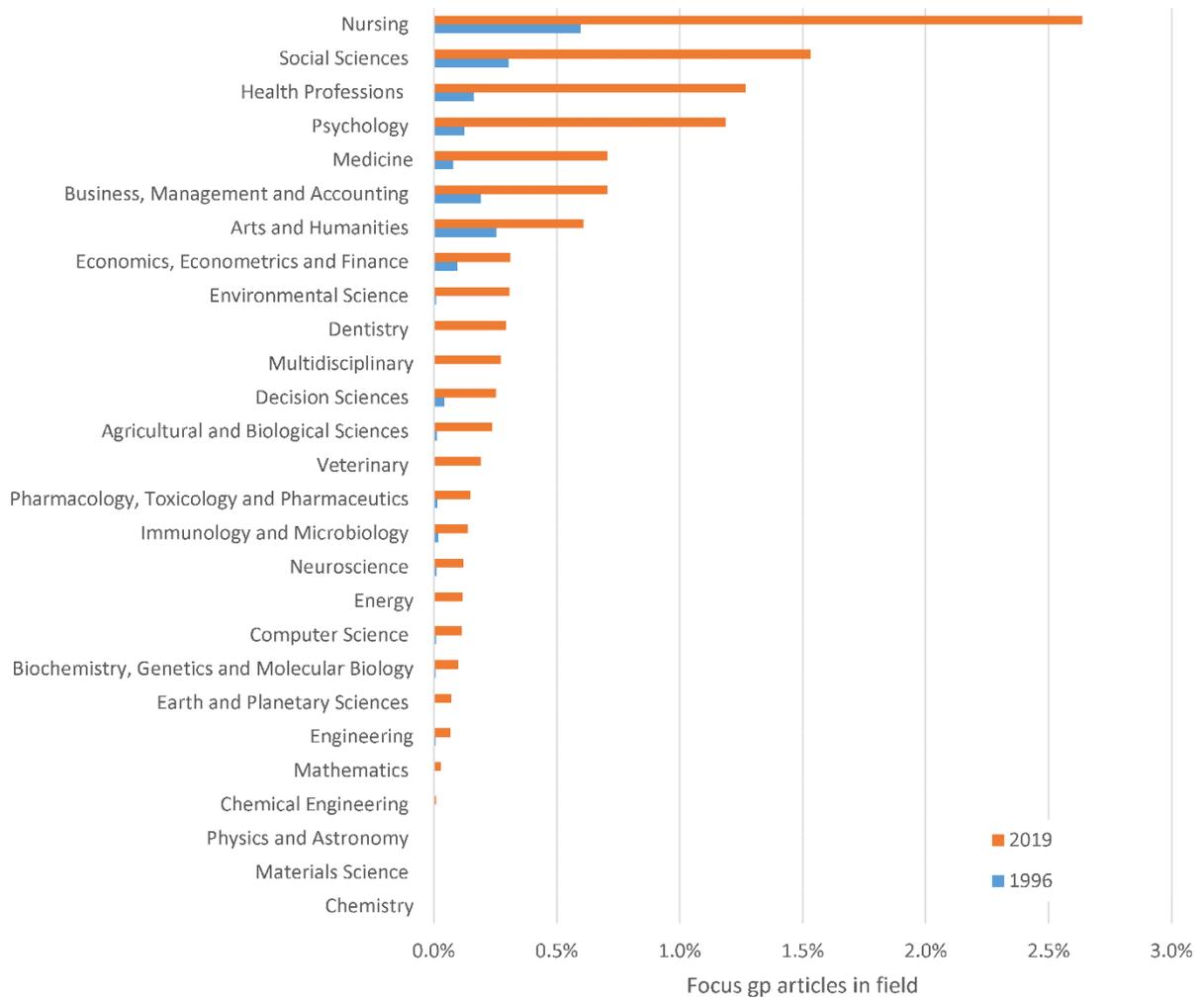

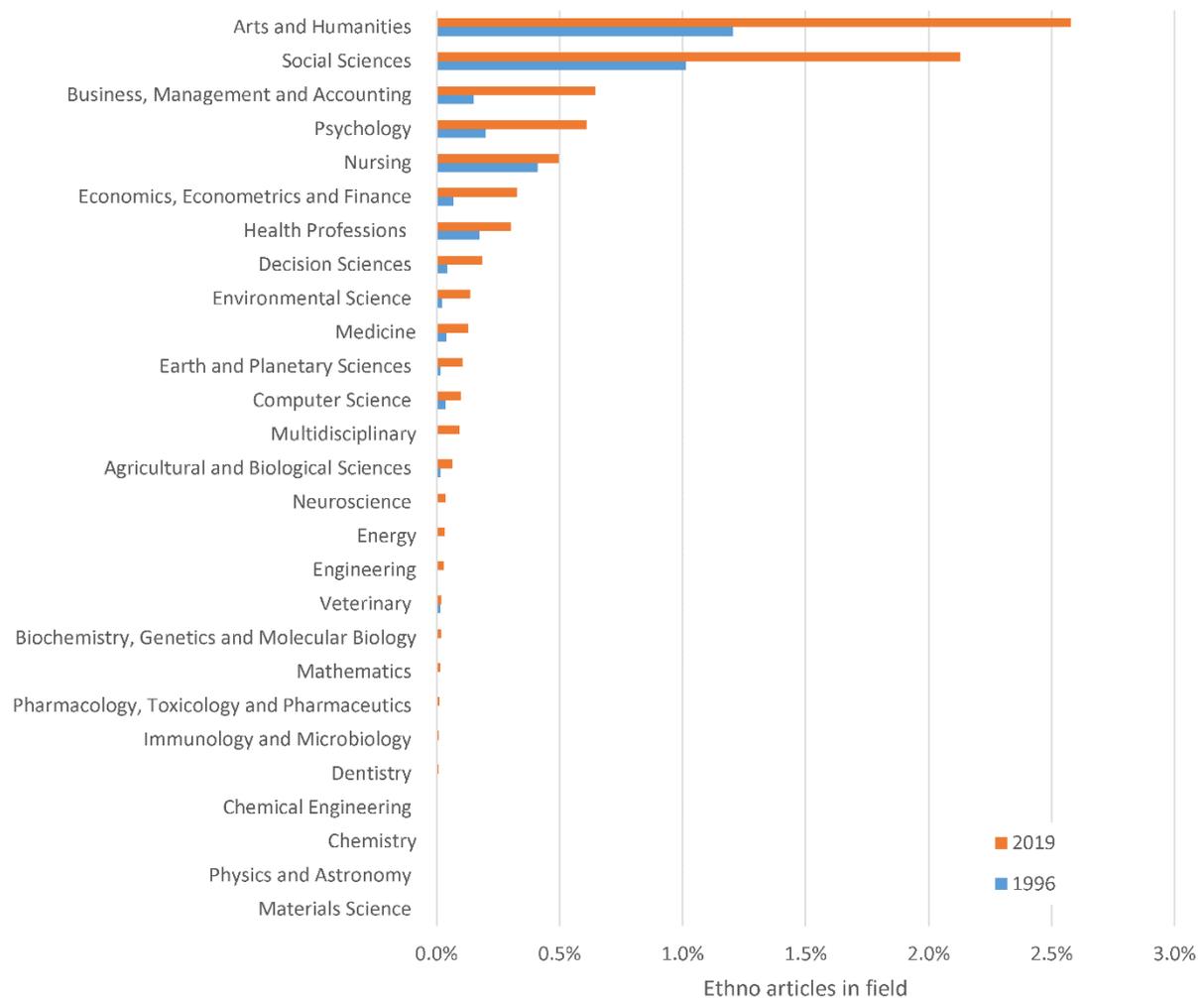

Figure 1. The proportion of Scopus journal articles mentioning *interviews, case studies, focus groups* or *ethnography* in their titles, abstract or keywords in 1996 and 2019 (qualification: at least 500 characters in the abstract).

Comparing the prevalence of the different methods between broad fields may help to identify patterns in the results (Figure 2). There is little relationship between the prevalence of interviews and case studies, except that they are both rare in a third of fields. It seems likely that fields using case studies much more often than interviews (e.g., Decision Sciences, Environmental Science, Energy, Earth and Planetary Sciences) would use the term "case study" for quantitative evaluations of geographically-specific phenomena rather than in the named qualitative research method sense. Although interviews are four times as prevalent as focus groups, there is little variation between broad fields in the sense of an above average preference for one (Figure 2). In contrast, ethnography seems to be universally rare except in the Arts & Humanities and Social Sciences.

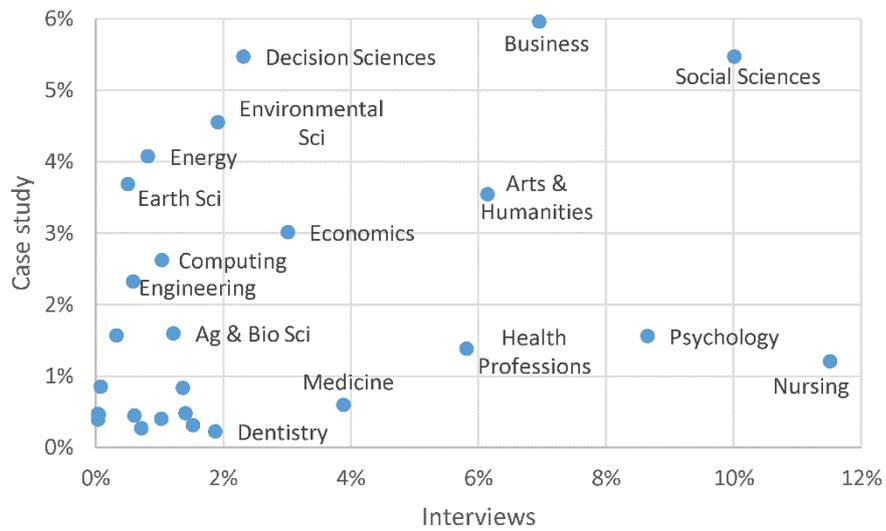
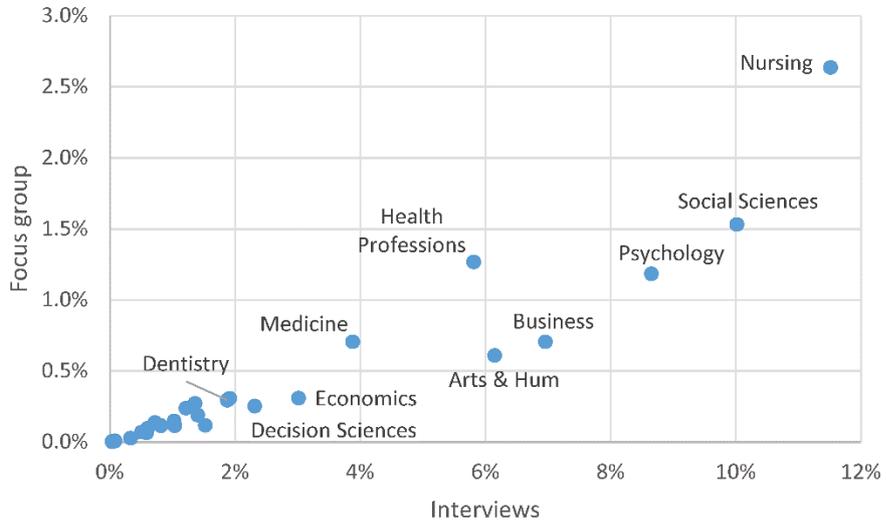
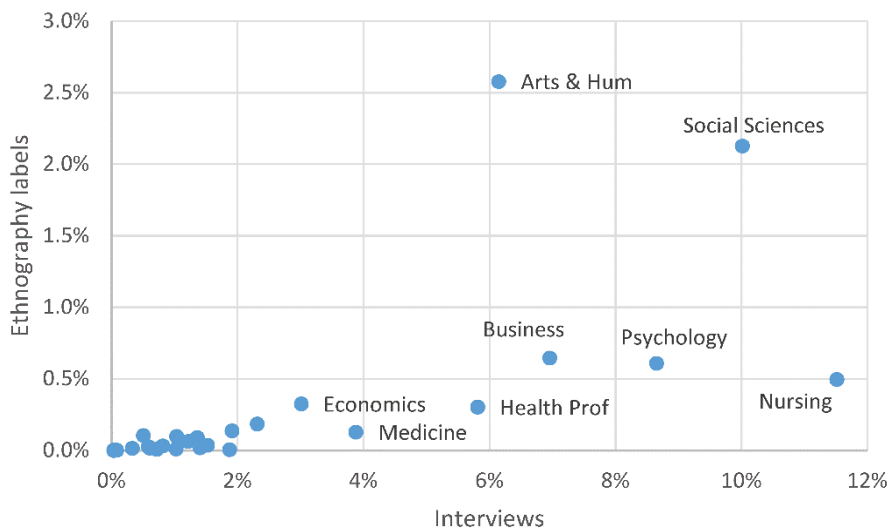

Figure 2. Percentages of case study, focus group and ethnography research against interview research in 2019.

*Impact*

The citation impact of interview research has mostly decreased over time, from being above average (1 is always the world average for the indicator used) in 1996 to below average in 2019 (Figure 3). Focus groups followed a similar pattern. In contrast, case study research has increased from being mostly below average in 1996 to about average overall in 2019, and above average in ten fields. The average citation impact of ethnographies is variable overall due to low numbers, but was above average in both 1996 and 2019 in its two core broad fields, Social Sciences and Arts & Humanities.

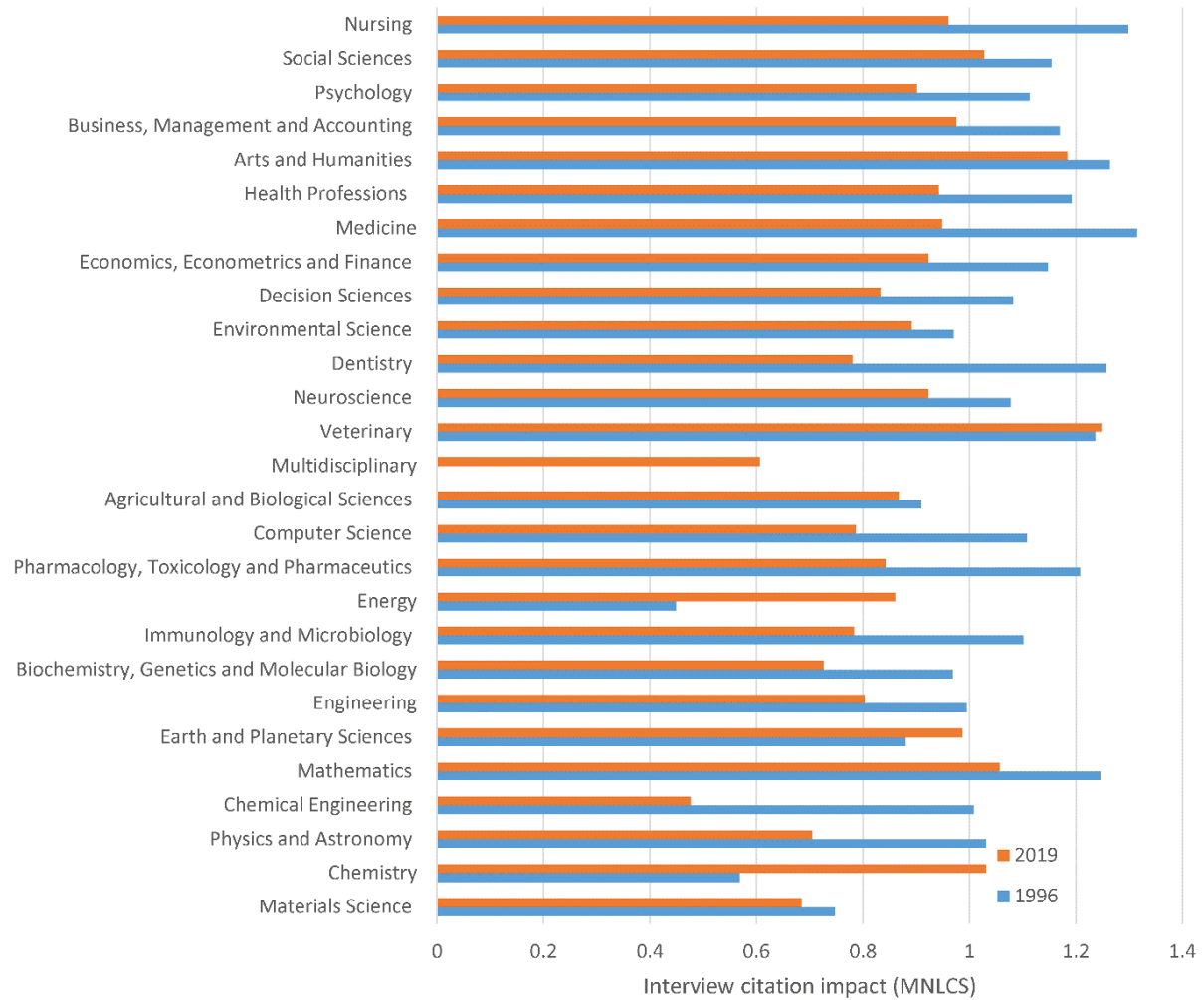

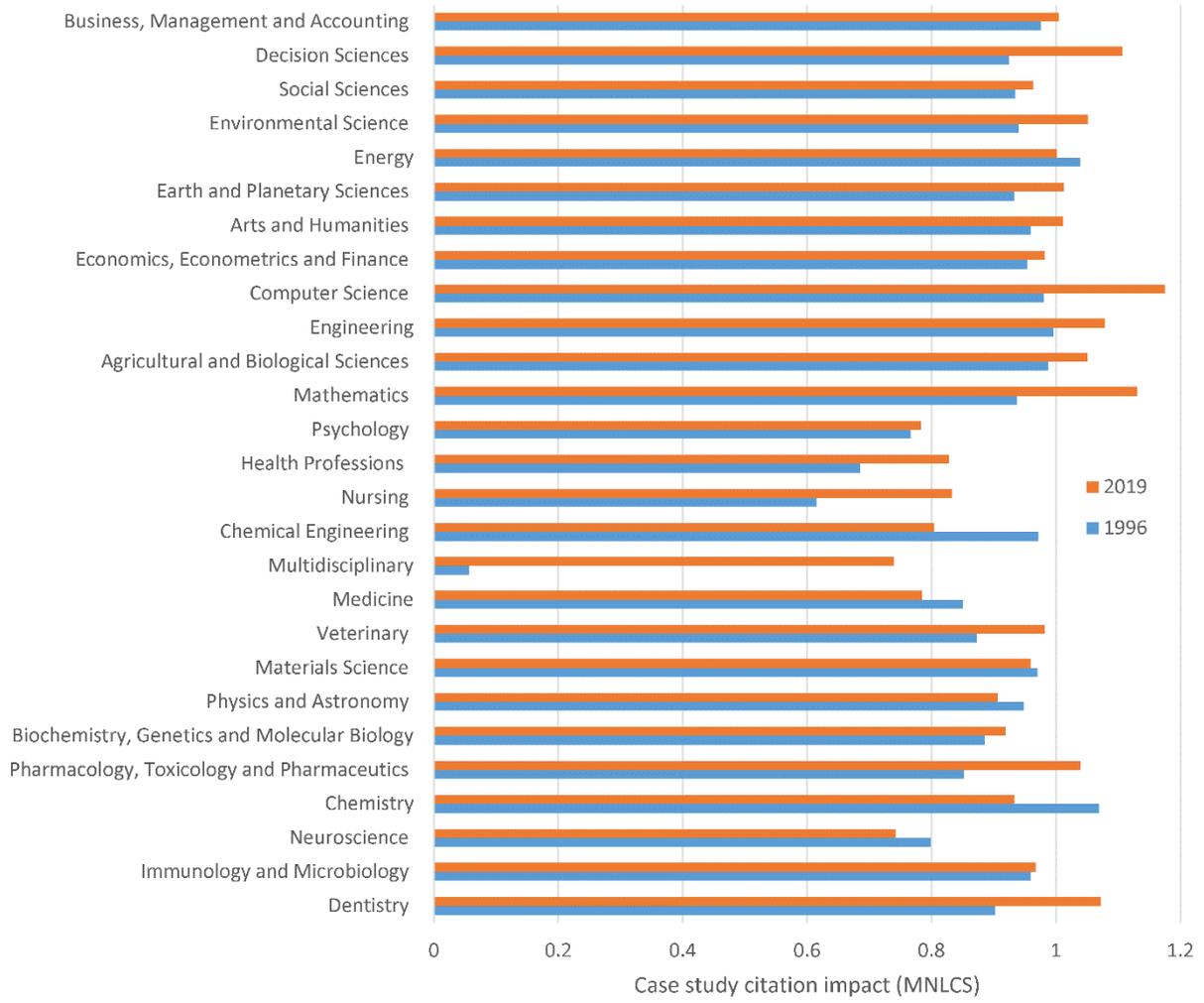

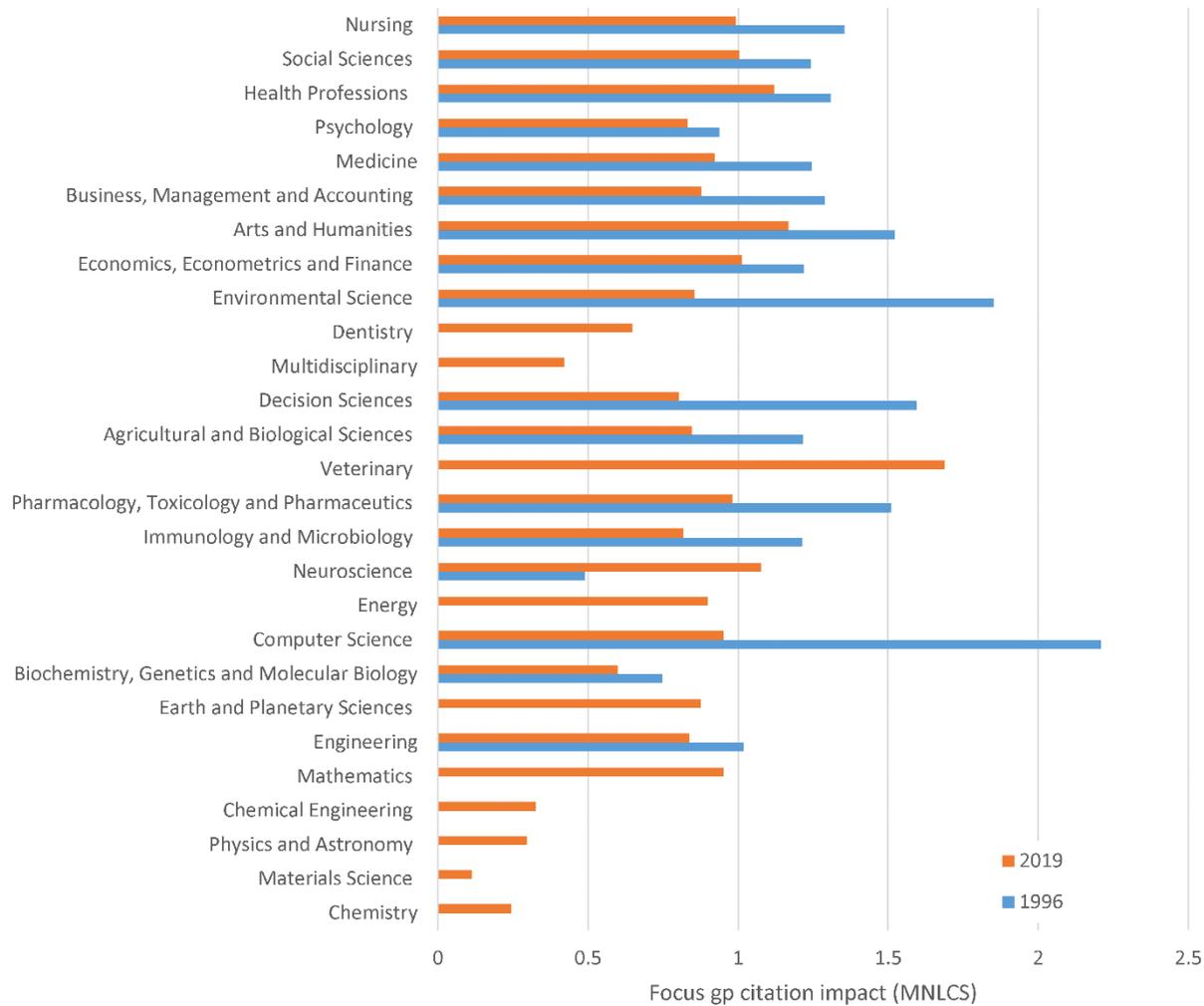

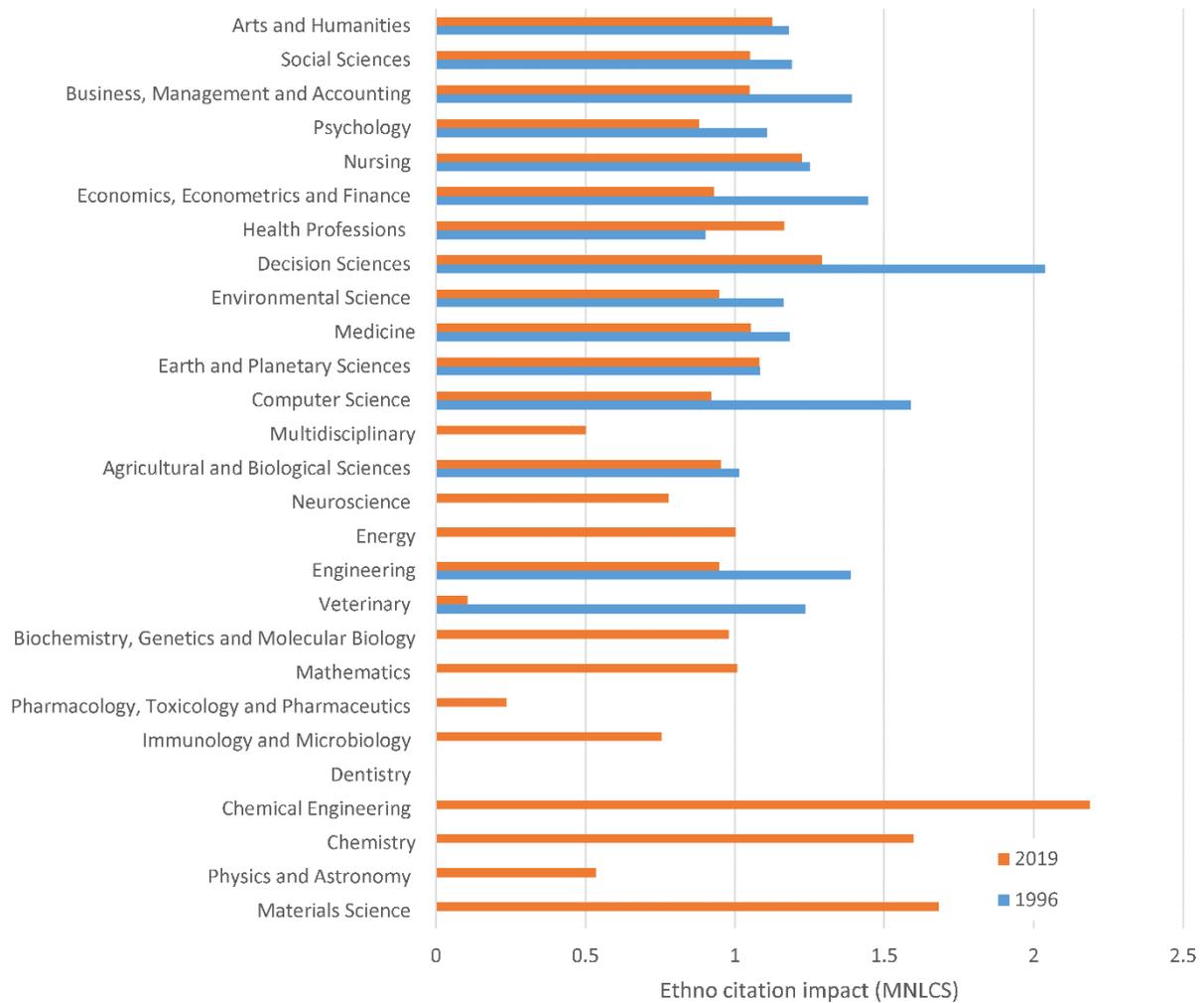

Figure 3. The average citation impact (MNLCS) of Scopus journal articles mentioning *interviews, case studies, focus groups* or *ethnography* in their titles, abstract or keywords in 1996 and 2019 (qualification: at least 500 characters in the abstract). Broad field MNLCS have been adjusted to equal 1. Fields are ordered as in Figure 1.

## Prevalence vs Impact

There is a general, but weak tendency for the citation impact of each method to be higher in fields where the method is more prevalent (i.e., a positive correlation for all years in Figure 4). This tendency has generally increased between 1996 and 2019 for interviews and case studies, but not for focus groups and ethnographies.

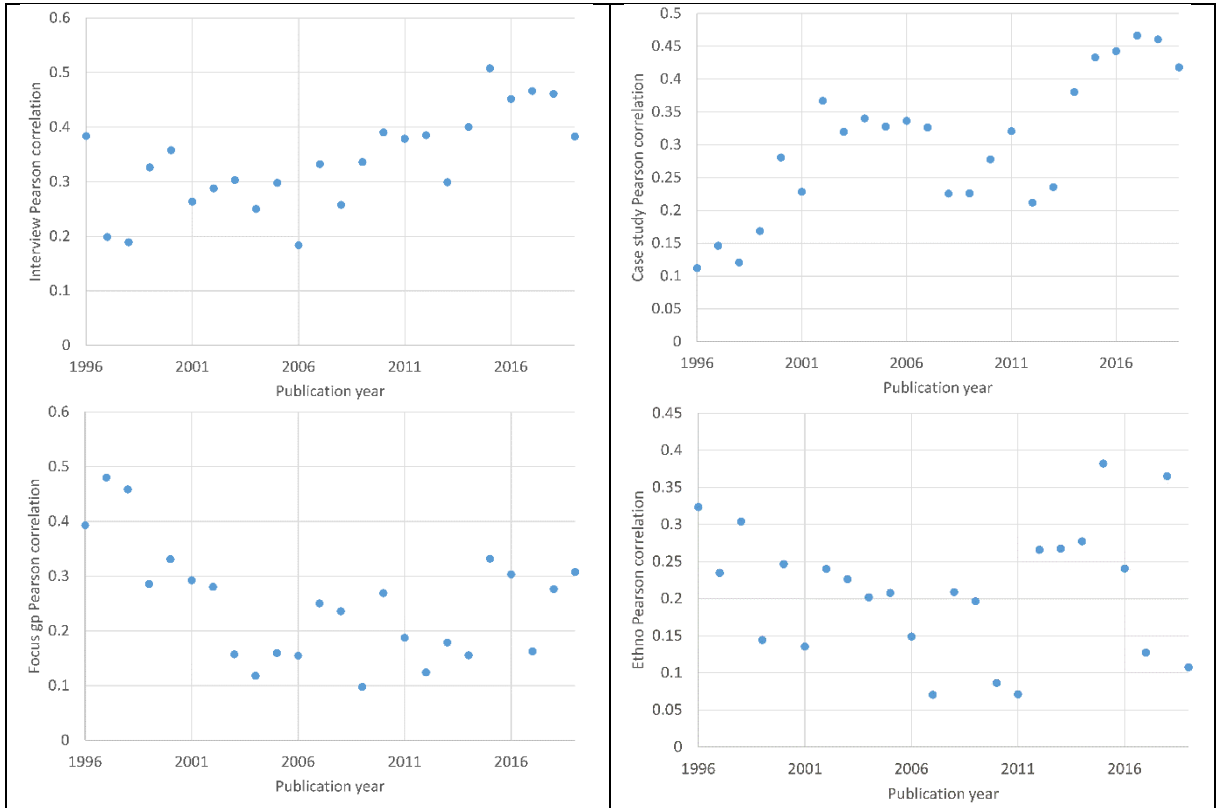

Figure 4. Pearson correlations between the proportion of Scopus journal articles mentioning *interviews, case studies, focus groups* or *ethnography* in their titles, abstract or keywords and their MNLCS (qualification: at least 500 characters in the abstract).

## Individual broad fields

The results above mostly contrast 2019 to 1996, but year-on-year changes are not always smooth (Figure 5, 6, 7, 8). Sharp jumps between years can occur as a side-effect of journals being added or removed from the database (e.g., Nursing interviews) or due to natural variation when small numbers are involved (e.g., Arts & Humanities focus groups).

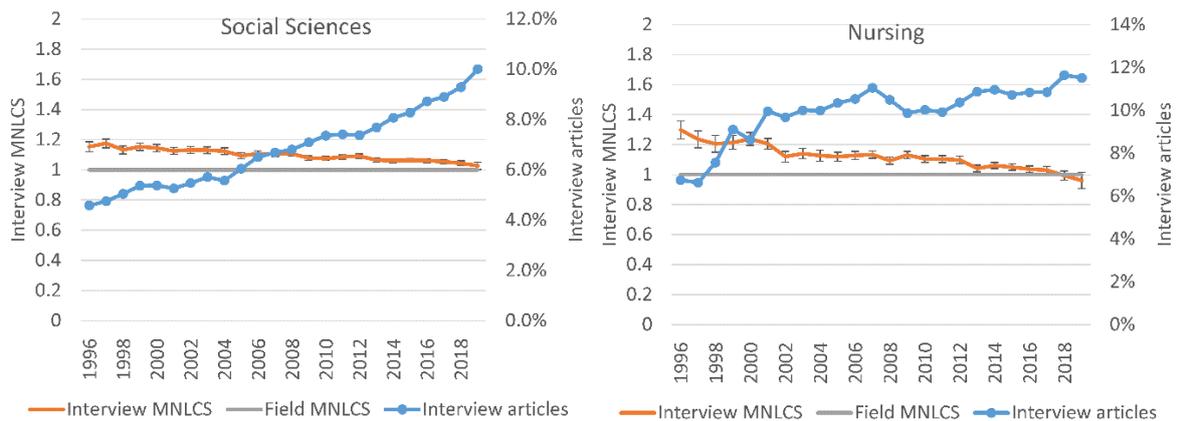

Figure 5. The proportion and MNLCS for Scopus journal articles mentioning *interviews* in their titles, abstract or keywords in 1996 and 2019 for two broad fields (qualification: at least 500 characters in the abstract). Graphs for the remaining 25 broad fields are in the online supplement. Error bars indicate 95% MNLCS confidence intervals.

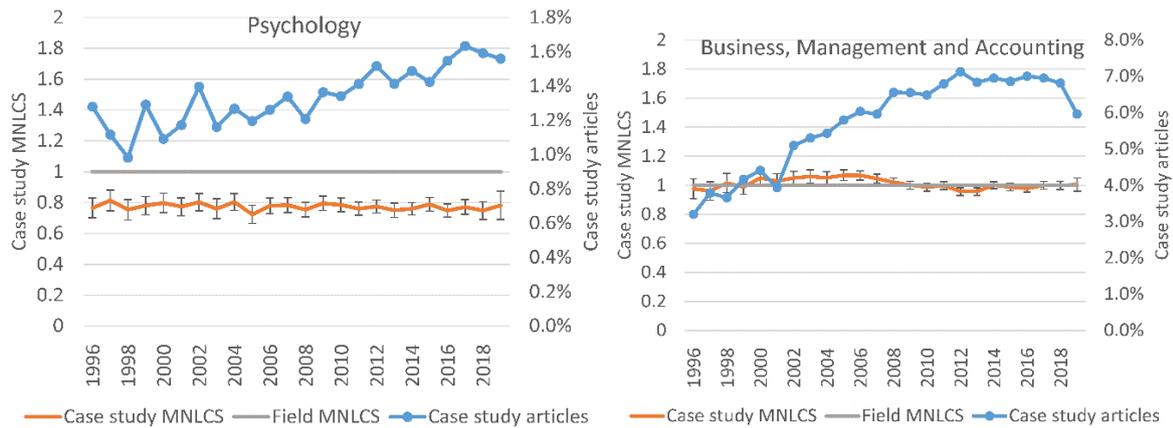

Figure 6. The proportion and MNLCS for Scopus journal articles mentioning *case studies* in their titles, abstract or keywords in 1996 and 2019 for two broad fields (qualification: at least 500 characters in the abstract). Graphs for the remaining 25 broad fields are in the online supplement. Error bars indicate 95% MNLCS confidence intervals.

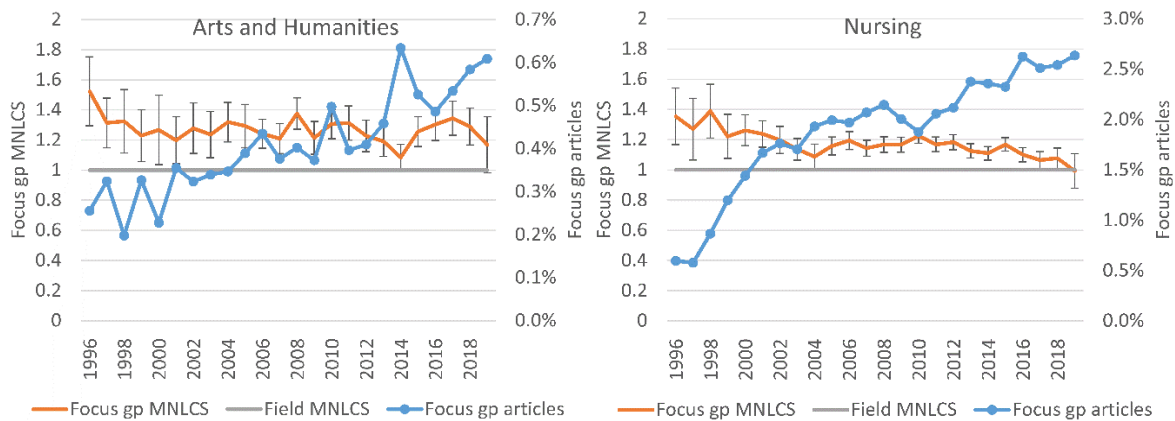

Figure 7. The proportion and MNLCS for Scopus journal articles mentioning *focus groups* in their titles, abstract or keywords in 1996 and 2019 for two broad fields (qualification: at least 500 characters in the abstract). Graphs for the remaining 25 broad fields are in the online supplement. Error bars indicate 95% MNLCS confidence intervals.

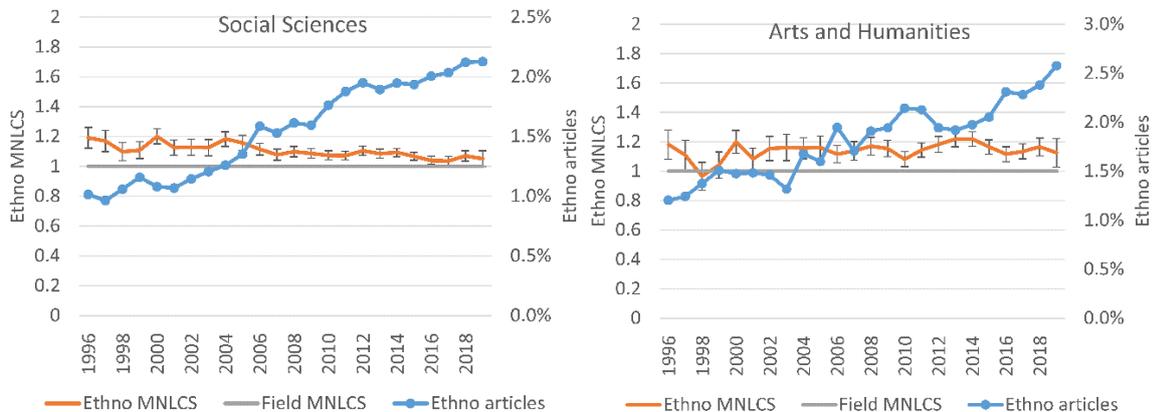

Figure 8. The proportion and MNLCS for Scopus journal articles mentioning *ethnography* in their titles, abstract or keywords in 1996 and 2019 for two broad fields (qualification: at least 500 characters in the abstract). Graphs for the remaining 25 broad fields are in the online supplement. Error bars indicate 95% MNLCS confidence intervals.

# Discussion

## *Limitations*

As mentioned above, a limitation of the method used here is that all the terms investigated (perhaps case studies in particular) may be used without the article analysing any qualitative data (i.e., a proportion of false results). In addition, the techniques may be used without explicitly being named in an article's title, abstract or author keywords (i.e., incomplete results). Another limitation is that the focus on journal articles excludes important research output types, including monographs or edited books. It is possible that this has influenced the results. If ethnographies are more likely to be written up as monographs, then they would be more prevalent than the graphs suggest, for example. These are all typical issues with quantitative research: the need to make simplifications in order to get data to analyse.

## *Comparison with prior research*

The results suggest that the increases in qualitative methods previously found for narrow fields ((Bluhm, Harman, Lee, & Mitchell, 2011;Hider, & Pymm, 2008;Taipale & Fortunati, 2014) are almost universal throughout broad fields, and apply to four individual approaches: interviews, case studies, focus groups and ethnographies. The exceptions previously found (Mullen, Budeva, & Doney, 2009; Wells, Kolek, Williams, & Saunders, 2015) are therefore rare exceptions in academia. The results also conflict with prior claims that big data approaches will overtake sociology, threatening traditional methods (Savage & Burrows, 2009), although sociology was not assessed on its own.

The analysis here has not diagnosed the cause of the increase in qualitative approaches, but prior findings of increases in mixed methods research (Alasuutari, 2010; Draper et al., 2018; Nunkoo et al., 2020) and a decline in conceptual research (Alasuutari, 2010; Nunkoo et al., 2020) suggests two potentially contributing factors.

## *Interviews: The standard qualitative data gathering approach?*

Interviews seem to be by far the most prevalent method for obtaining qualitative data, with close to 12% of Nursing articles in 2019 mentioning interviews in their titles, abstracts or author keywords. Interviews are twice as prevalent as case studies, and probably more considering the false matches for case studies (see below). They are four times as prevalent as the next most reliable method, focus groups. Interviews are natural methods to obtain data for people-focused subjects. Nursing research may be particularly reliant on interviews because it presumably focuses on nurses or patients rather than equipment or technology. Focus groups may be substantially less useful because of the ethical overheads associated with patient focus groups.

Interviews are used to some extent in all broad fields but are extremely rare in the physical sciences. To illustrate how interviews occur in the rarest category, Materials Science, one matching article from 2019 was, "Identification of problems arising during manual handling of food packaging by older consumers in Poland" from the journal, *Packaging Technology and Science* (Świda, Halagarda, Prusak, & Popek, 2019). This article reported in-depth interviews with consumers, clearly generating qualitative data. It deals with the consumer side of materials science rather than the physical science side and is therefore application focused, but a valid contribution to Materials Science. This confirms that the term "interview" is used in the context of a qualitative analysis. There were also a few transcripts of interviews with prominent individuals, such as "An interview with David Foord, Director of Materials Science at Thermo Fisher Scientific" (McGuire, & Foord, 2019), which reported a traditional journalistic set of questions and answers rather than a qualitative study. This is not an academic journal article but was classified as such in Scopus.

### *Case studies: Evidence of localisation and/or qualitative approaches?*

The use of the term case study was checked for a qualitative connotation in its most prevalent domain, Business, Management, and Accounting, by reading a selection of articles from 2019 matching the query. The term appeared to be widely used in a non-qualitative context. For example, two of the first four matches implied this, "…quantitatively through simulation. Simulation was demonstrated through a single case study…" and "…computed the value of the marginal productivity of different inputs in three selected case studies (Angola, Mozambique and Brazil).". Thus, the term "case study" often refers to localised circumstances rather than a qualitative methodology. The prevalence of the term within some broad fields may reflect the extent to which analysing general phenomena in local or narrow contexts is relevant to knowledge creation.

### *Focus groups: Complementing interviews, but not for senior roles?*

Focus groups are used proportionately to interviews in most fields and presumably for similar reasons. There is a slight tendency for health-related fields to use focus groups proportionately more, however, although it is not clear why. Focus groups are proportionately rarer in the Arts and Humanities and Business, Management and Accounting compared to interviews. In these cases, some of the interviews might be with people in senior or creative roles (e.g., managers, artists), and focus groups would be inappropriate for this type of application. For example, one Arts and Humanities study interviewed heads of libraries in Ghana (Dzandza, 2019) and another conducted "interviews with stakeholders" from the cultural industry in Syria (Kousa & Pottgiesser, 2019). In a similar Business, Management and Accounting example, "18 semi-structured interviews were carried out with chief investigators" (DeSisto, Cavanagh, & Bartram, 2019).

### *Ethnography: Just for the social sciences, arts and humanities?*

Ethnography is mentioned at the highest rate in the Arts & Humanities and Social Sciences, presumably to get insights into localised social or cultural contexts, especially those that are in some senses remote to the researcher. Ethnography is comparatively common in the Economics, Econometrics and Finance broad field, partly because of multidisciplinary journals overlapping with business and marketing (e.g., consumer culture articles), although there is also a Research in Economic Anthropology edited book series (classified as a journal by Scopus), which includes ethnographies of groups engaging in economic activities, such as, "people who consume and sell fair trade in the Italian city of Palermo" (Orlando, 2019).

Ethnography is rare compared to interviews in health-related fields, perhaps because health professionals conduct the research in these areas and have regular contact with patients and each other. This may largely negate the value of participant-observation ethnographic approaches, at least in clinical settings. Moreover, the ethical drawbacks of participant-observation ethnographies amongst patient groups would presumably be nontrivial in some cases. Exceptions might include ethnographies of sufferer-managed support groups to investigate their efficacy (e.g., Pereira Neto, Barbosa, Silva, & Dantas, 2015).

# Conclusions

The results suggest that qualitative research increased substantially in importance between 1996 and 2019. As apparently the first systematic evidence throughout academia of the prevalence of qualitative research, this should reassure practitioners, editors and publishers. Arguments that some qualitative methods are threatened by big data are not supported by the results, although this may still occur within some narrow specialisms. A tendency for the citation impact of interview-based research to decline is the main concern raised by the results, especially because it was below average in most fields in 2019 (exceptions include Social Sciences and Arts & Humanities). Although citation impact may well be less important than societal impact for interview-based research, this suggests that extra care should be given to checking the quality of interview-based research in case quality issues have caused the decline. Some journals identifying problems with particular methods have published guideline articles to help authors (e.g., Hartel, 2020), and this would be an appropriate response, if warranted.

Finally, whilst qualitative research understandably remains a minority pursuit in many broad fields, it is increasingly accepted in all. Thus, editors, reviewers and methods teachers in the era of big data should not neglect the core qualitative approaches. Moreover, they should be considered as reasonable and increasingly mainstream approaches in all areas of scholarship.